\documentclass[11pt]{cernrep}
\usepackage{graphicx}
\usepackage{here}
\begin{document}
 \title{Uncertainties on parton distribution functions from the ZEUS 
NLO QCD fit to data on deep inelastic scattering}

\author{A.M.Cooper-Sarkar}
\institute{Particle Physics, University of Oxford, UK}
\maketitle
\begin{abstract}
   An NLO QCD analysis of the ZEUS data on $e^+ p$
   deep inelastic scattering 
   together with fixed-target data has been performed from which 
   the gluon and quark densities of the proton and the value of the strong
   coupling parameter, 
   $\alpha_s(M_Z^2)$, have been extracted. The study includes a full
   treatment of the experimental systematic uncertainties, 
   including point-to-point correlations. Different ways of
   incorporating correlated systematic uncertainties into the fit
   are discussed and compared.
   \end{abstract}

\section{INTRODUCTION}

Studies of inclusive differential cross sections and structure functions, 
as measured in deep inelastic scattering (DIS) of leptons from hadron
targets, have played a crucial role in
establishing the theory of perturbative quantum chromodynamics (pQCD).
Measurement of the structure functions as a
function of $x$ and $Q^2$ yields information on the shape 
 of the parton distribution functions (PDFs) and, 
through their $Q^2$ dependence, on the value
of the strong coupling constant $\alpha_s(M_Z^2)$. Most analyses use the
formalism of the next-to-leading-order (NLO) DGLAP evolution 
equations~\cite{1}
which provide a successful description 
of the data over a broad kinematic range.

In recent years the uncertainties on PDFs from experimental sources, as well
as from model assumptions, have become an 
issue. The subject of this paper is an evaluation of the experimental 
uncertainties on the extracted PDFs and on the value of $\alpha_s(M_Z^2)$.
Various methods of treating correlated systematic uncertainties are discussed.
The method selected for the main analysis 
is conservative, reflecting knowledge that
such systematic uncertainties are not always Gaussian distributed.
Model uncertainties have also been estimated.

\section{Description of NLOQCD fit}

Full details of the analysis are given in~\cite{2}, 
here we give only a summary. The kinematics
of lepton hadron scattering is described in terms of the variables $Q^2$, the
negative invariant mass squared 
of the exchanged vector boson, Bjorken $x$, the fraction
of the momentum of the incoming nucleon taken by the struck quark (in the 
quark-parton model), and $y$, which measures the energy transfer between the
lepton and hadron systems.
The differential cross-section for the process $e^+ p \to e^+ X$ 
is given in terms of the structure functions by
\[
\frac {d^2\sigma } {dxdQ^2} =  \frac {2\pi\alpha^2} {Q^4 x}
\left[Y_+\,F_2(x,Q^2) - y^2 \,F_L(x,Q^2)
- Y_-\, xF_3(x,Q^2) \right],
\]
where $\displaystyle Y_\pm=1\pm(1-y)^2$. 
The structure functions are directly related to PDFs, 
and their $Q^2$ dependence, or scaling violation, 
is predicted by pQCD. For $Q^2 \le 1000$GeV$^2$, $F_2$ dominates the
charged lepton-hadron cross-section and for $x \le 10^{-2}$ 
the gluon contribution
dominates the $Q^2$ evolution of $F_2$, such that HERA data in this kinematic
region provide crucial information
on quark and gluon distributions. (Schematically, $F_2 \sim xq$,
 $dF_2/dlnQ^2 \sim \alpha_s P_{qg} xg$).

A global fit of ZEUS~\cite{3} and fixed target DIS data{\cite{4} 
has been perfomed.
The fixed target data are used to provide information on the
valence quark distributions and the flavour composition of the sea, 
and to constrain the fits at high $x$. 
All data sets used have
full information on point-to-point correlated systematic uncertainties. 
(ZEUS $e^+p$ cross-section data; NMC, E665, BCDMS $\mu-p$ and $\mu-D$ $F_2$ 
data; CCFR $\nu,\bar{\nu}$ $xF_3$ data on an Fe target).
A total of 71 sources of systematic uncertainty, including normalisation
uncertainties, were included.

The analysis is performed within the conventional
framework of leading twist, NLO QCD, with the
renormalisation and factorization scales chosen to be $Q^2$. 
In the standard fit the following cuts are made on
the ZEUS and the fixed target data: (i)~$W^2 > 20$GeV$^2$ to reduce the
sensitivity to target mass and higher 
twist contributions, which become important at high $x$
and low $Q^2$; (ii)~$Q^2 > 2.5$GeV$^2$ to remain in the kinematic region where
perturbative QCD should be applicable. 
The heavy quark production scheme is the
general mass variable flavour number scheme of Thorne and Roberts~\cite{5}.

The DGLAP equations yield the PDFs at all values of $Q^2$, provided they
are input as functions of $x$ at some input scale $Q^2_0$.
The PDFs for $u$ valence, $d$ valence, total sea ($xS$), gluon ($xg$) and 
the difference between the $d$ and $u$
contributions to the sea, are each parametrized  by the form 
\[
   xf(x) =  p_1 x^{p_2} (1-x)^{p_3}( 1 + p_5 x)
\]
at $Q^2_0 = 7$GeV$^2$. The flavour structure of the light quark sea 
allows for the violation of the Gottfried sum rule and the
strange sea is suppressed by a factor of 2 at $Q^2_0$,
consistent with neutrino induced dimuon data from CCFR. 
The parameters $p_1-p_5$ are constrained to impose the momentum sum-rule and 
the number sum-rules
on the valence distributions. There are 11 
free parameters in the standard fit when the strong coupling constant
 is fixed to $\alpha_s(M_Z^2) =  0.118$~\cite{6},
and 12 free parameters when $\alpha_s(M_Z^2)$ is 
determined by the fit.

\section{Definition of $\chi^2$: treatment of correlated systematic uncertainties}
\label{sec:errors}

The definition of $\chi^2$ used in global fits 
has traditionally been
\[
\chi^2 = \sum_i \frac{\left[ F_i^{\rm NLOQCD}(p)-F_i(\rm meas) \right]^2}{(\sigma_{i,\rm stat}^2+\sigma_{i,\rm unc}^2+\sigma_{i,\rm corr}^2)} 
\]
where $F_i^{\rm NLOQCD}(p)$ represents the prediction 
from NLO QCD in terms of the theoretical parameters $p$; 
$F_i(\rm meas)$ represents a measured data point
and the symbols $\sigma_{i,\rm stat}$, $\sigma_{i,\rm unc}$ and
$\sigma_{i,\rm corr}$ represent its error from statistical, uncorrelated 
and correlated systematic sources,
respectively.

However such a definition does not take into account the correlations of the
correlated systematic errors. Hence it has been modified to
\[
\chi^2 = \sum_i \frac{\left[ F_i(p,s)-F_i(\rm meas) \right]^2}{(\sigma_{i,\rm stat}^2+\sigma_{i,\rm unc}^2)} + \sum_\lambda s^2_\lambda 
\label{eq:chi2}
\]
where
\begin{equation}
F_i(p,s) = F_i^{\rm NLOQCD}(p) + 
\sum_{\lambda} s_{\lambda} \Delta^{\rm sys}_{i\lambda}
\label{eq:predmod}
\end{equation}
Eq.(\ref{eq:predmod}) shows how the theoretical prediction is modified
to include the effect of the correlated systematic uncertainties. 
The one-standard-deviation systematic uncertainty 
on data point $i$ due to source $\lambda$ is
referred to as $\Delta^{\rm sys}_{i\lambda}$ and 
the parameters $s_\lambda$ represent independent Gaussian random variables 
with zero mean and unit variance for each source of
 systematic uncertainty. 
There are then several different ways to proceed, as discussed below.
 
\subsection{Offset methods}
The systematic uncertainty parameters $s_\lambda$ can be fixed to zero 
so that the fitted theoretical predictions are as close as 
possible to the central values of the published data.
However, the $s_\lambda$ are allowed  to vary for the error analysis, 
such that in addition to 
the usual Hessian matrix, $M_{jk}$, given by
\[
M_{jk}= \frac{1}{2}\frac{\partial^2 \chi^2}{\partial p_j \partial p_k},
\]
which is evaluated with respect to the theoretical parameters,
a second Hessian matrix, $C_{j\lambda}$, given by
\[
C_{j\lambda} = \frac{1}{2}\frac{\partial^2 \chi^2}{\partial p_j \partial s_{\lambda}}
\]
is evaluated. The systematic covariance matrix is then given by
$V^{ps}= M^{-1} C C^T M^{-1}$~\cite{7} and the total
covariance matrix by  
$V^{\rm tot} = V^p +V^{ps}$, where $V^p=M^{-1}$. 
Then the uncertainty on any distribution $F$ may be calculated from
\[
<\Delta F^2>=\sum_j \sum_k \frac{\partial F}{\partial p_j} V_{jk} \frac{\partial F}{\partial p_k} 
\]
by substituting $V^p$, $V^{ps}$ or $V^{\rm tot}$ for $V$, to obtain the
statistical (and uncorrelated systematic), correlated systematic 
or total experimental error band, respectively. 

This method of accounting for systematic uncertainties will be called the
`offset method' since its results are equivalent to those of
the traditional offset method used by experimentalists, in which
each $s_\lambda$ is varied by its assumed uncertainty ($\pm1$) (such that the
data points are shifted to account for systematic error $\lambda$) 
a new fit is performed for each of these variations,
and the resulting deviations of the theoretical parameters 
from their  central  values are added in 
quadrature~\cite{8}. (Positive and  negative deviations are added 
in quadrature separately). 
This is not a statistically rigorous procedure, but its virtue is that
it does not assume that the systematic errors are
necessarily Gaussian distributed. It gives a conservative estimate of the 
error as compared to the Hessian methods~\cite{8,9}, which will be 
described below.

\subsection{Hessian methods}

An alternative procedure would be to allow the systematic
uncertainty parameters $s_\lambda$ to vary in the main fit 
when determining the values of the
theoretical parameters. 
This method is referred to as `Hessian method 1'.
The errors on the theoretical parameters are then calculated from the
inverse of a single Hessian matrix which expresses the variation of 
$\chi^2$ with respect to both 
theoretical and systematic offset parameters.
Effectively, the theoretical prediction is not fitted
to the central values of the published experimental data, but allows 
these data points to move within the tolerance of their correlated
systematic uncertainties. 
It is necessary to check that
points are not moved far outside their one standard deviation systematic 
uncertainty estimates. The theoretical prediction determines the 
optimal settings for correlated systematic shifts of experimental data points 
such that the most consistent fit to all data sets is obtained. Thus
systematic shifts in 
one experiment are correlated to those in another experiment by the fit.
 
Hessian method 1 becomes
an impractical procedure when the number of sources of systematic uncertainty
is large, as in  the present global DIS analysis in which 71 independent 
sources of systematic uncertainty were included. Recently
CTEQ~\cite{10} have given an elegant analytic method for
performing the minimization with respect to systematic-uncertainty parameters.
This gives a new formulation of the $\chi^2$

\[
\chi^2 = \sum_i \frac{\left[ F_i^{\rm NLOQCD}(p)-F_i(\rm meas) \right]^2}{(\sigma_{i,\rm stat}^2+\sigma_{i,\rm unc}^2)}  - B A^{-1} B
\]
where
\[
B_\lambda = \sum_i \Delta_{i\lambda}^{\rm sys} \frac{\left[F_i^{\rm NLOQCD}(p)-F_i(\rm meas)\right]}{(\sigma_{i,\rm stat}^2+\sigma_{i,\rm unc}^2)}
\]
and
\[
A_{\lambda\nu} = \delta_{\lambda\nu} + \sum_i \Delta_{i\lambda}^{\rm sys} \Delta_{i\nu}^{\rm sys} /(\sigma_{i,\rm stat}^2+\sigma_{i,\rm unc}^2),
\]
such that the uncorrelated and systematic 
contributions to the $\chi^2$ can be evaluated separately.
This method is referred to as `Hessian method 2'.

The results for the ZEUS fit analysis are compared for these methods below.

\section{Fit results: experimental and model uncertainties} 
\subsection{Experimental undertainties: offset method}
\label{sec:standard}

The standard fit has been perfomed treating the experimental 
correlated systematic errors by the offset method, 
with $\alpha_s(M_Z^2) = 0.118$ fixed.
The fit gives an excellent description of the high-precision ZEUS data
and of the fixed-target data, see ref~\cite{3}.
The sea and the gluon PDFs extracted from this fit are shown in Fig.~\ref{fig:glusea}.
\begin{figure}
\includegraphics[scale=.65]{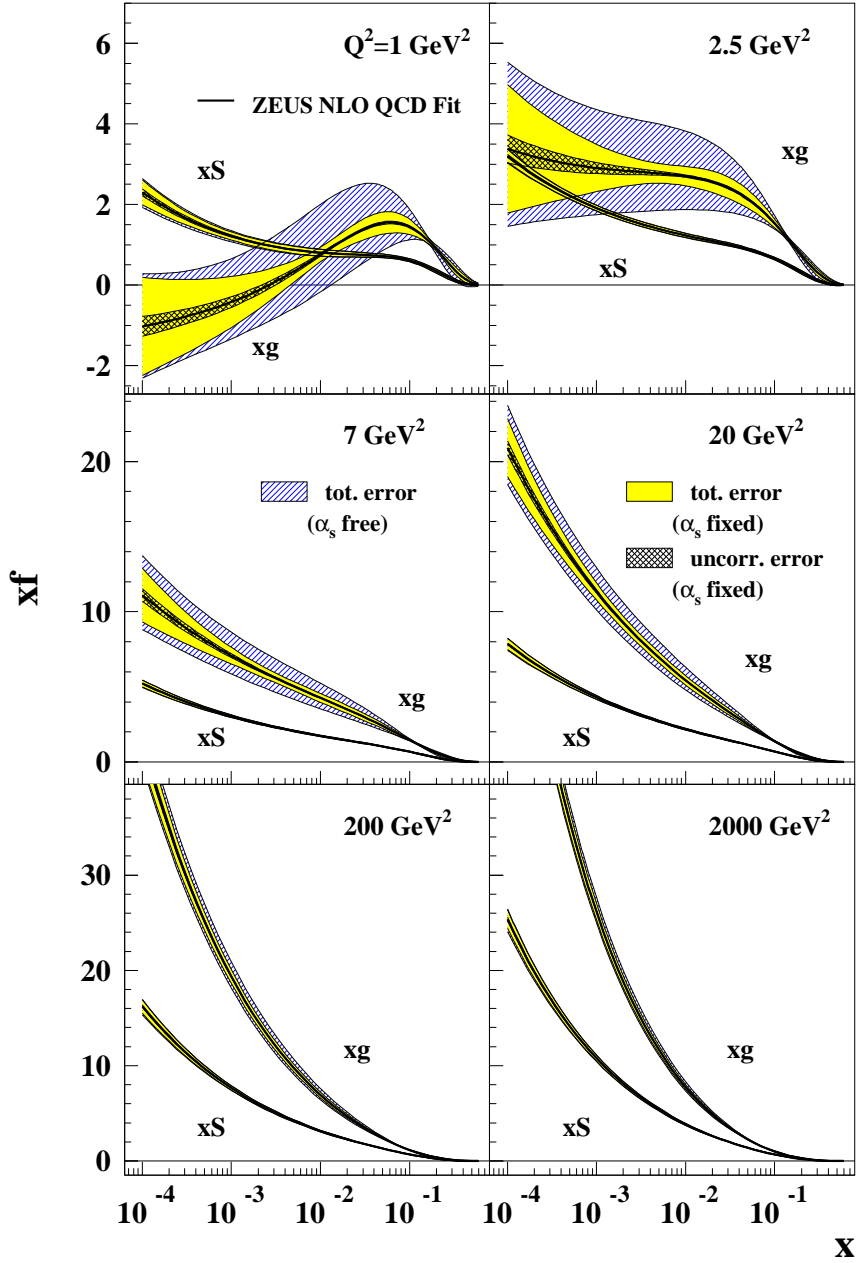}
\caption{Comparison of the  gluon and sea distributions from the 
ZEUS-S NLO QCD fit for various $Q^2$ values.
In this figure, the cross-hatched error bands show the statistical and 
uncorrelated systematic uncertainty, the grey error bands show the total 
experimental uncertainty including correlated systematic uncertainties 
(both evaluated from the standard fit with $\alpha_s(M_Z^2)=0.118$) 
and the hatched error bands show 
the additional uncertainty coming from variation 
of the strong coupling constant $\alpha_s(M_Z^2)$.
}
\label{fig:glusea}
\end{figure}
These PDFs agree well with the latest distributions from MRST2001~\cite{11} 
and CTEQ6~\cite{12}. 
The error bands shown in the figure illustrate the experimental uncertainties
from i) statistical and uncorrelated systematic uncertainties alone; ii) total
experimental uncertainty including correlated systematic uncertainties; and
ii) additional uncertainty due to allowing $\alpha_s(M_Z^2)$ to be a parameter
of the fit.

Clearly, in the latter case, the fit also determines the value of 
$\alpha_s(M_Z^2)$, with its correlations to the PDF parameters fully accounted.
The value
\begin{equation}
\alpha_s(M_Z^2) = 0.1166 \pm 0.0008(\rm uncorr) \pm 0.0032 (\rm corr) \pm 0.0036(\rm norm)
\label{eq:alphas}
\end{equation}
is obtained, where the three uncertainties arise from the following: 
statistical and other uncorrelated sources; correlated systematic sources from 
all contributing experiments except that from their normalisations; 
the contribution from the latter normalisations. The contribution from
normalisation uncertainties is shown separately from the other systematic
uncertainties since, for many experiments, quoted normalisation uncertainties 
represent the limits of a box-shaped distribution
rather than the  standard deviation of a Gaussian distribution. An alternative
evaluation of normalisation uncertainty, which accounts for this, would reduce
this contribution to $\pm 0.0010$.

\subsection{Model uncertainties}

In addition to the experimental uncertainty on the fitted parameters,
there is potentially a model uncertainty due to the specific assumptions
made when setting up the NLOQCD fit. 
Sources of model uncertainty within the theoretical framework 
of leading-twist NLO QCD are:
the effect of varying the value of $Q^2_0$, and the minimum $Q^2$, 
 $x$ and $W^2$ of data entering the fit;
variation of  the form of the input PDF parameterisations; 
the choice of the heavy-quark production scheme. 
The sensitivity of the results to the variation of these input 
assumptions has been quantified in terms of the resulting variation in 
$\alpha_s(M_Z^2)$, since it is the most sensitive parameter. 
This leads to a model uncertainty in
$\alpha_s(M_Z^2)$, of $\Delta \alpha_s(M_Z^2) \sim \pm 0.0018$, 
considerably smaller than the errors from correlated systematic and 
normalisation uncertainties. The PDF parameters are much less sensitive to 
the model assumptions than $\alpha_s(M_Z^2)$. It follows that 
the error bands illustrated on the 
parton densities in Fig.~\ref{fig:glusea} represent 
reasonable estimates of the total uncertainties {\it within the 
theoretical framework of leading twist NLO QCD}. 

Sources of uncertainty due to the theoretical framework itself are considered
in the contribution of R.Thorne to this meeting.

\subsection{Comparison of Offset and Hessian methods}

First Hessian method 1 and Hessian method 2 
have been compared in a fit to ZEUS data alone,
for which the systematic uncertainties and relative normalisations are 
well understood. The results are very similar, 
as expected if the systematic uncertainties are Gaussian and the values
$\Delta^{sys}_{i\lambda}$ represent one standard deviation uncertainties.
However, if data sets from different experiments are used in the fit, these two
Hessian methods are only similar if normalisation uncertainties
are not included among the systematic uncertainties. 
Normalisation uncertainties 
are not always Gaussian and thus the analytic procedure
of Hessian method 2 is inappropriate for them.

The offset method has been compared to Hessian method 2
by performing the fit to global DIS data 
using Hessian method 2  to calculate the $\chi^2$.
Normalisation uncertainties were excluded and $\alpha_s(M_Z^2)$ was included 
as one of the theoretical parameters. 
This fit yields, $\alpha_s(M_Z^2) = 0.1120 \pm 0.0013$,  where the error 
represents the total experimental uncertainty from correlated and uncorrelated
sources, excluding normalisation uncertainties.
Thus this value should be compared with, 
$\alpha_s(M_Z^2) = 0.1166 \pm 0.0033$, 
evaluated  using the offset method, also excluding normalisation uncertainties
(see Eq.~\ref{eq:alphas}).  
Hessian method  2 gives a much reduced `optimal' error estimate for 
$\alpha_s(M_Z^2)$ and this is also the case for the PDF parameters.
The value of $\alpha_s(M_Z^2)$ is shifted from that obtained by the offset 
method. The PDF parameters are not affected
as strongly, their values are shifted by amounts which are 
well within the error estimates quoted for the offset method.

To compare the $\chi^2$ of the fits done using the offset method and 
Hessian method 2, it is necessary to 
use a common method of $\chi^2$ calculation.
Table~\ref{tab:chibab} presents the $\chi^2$ for the theoretical parameters
 obtained using these methods, re-evaluated by adding 
statistical and systematic errors in quadrature.
For both methods $\alpha_s(M_Z^2)$ has been included
among the theoretical parameters and normalisation uncertainties have not been
included among the systematic parameters. The total increase of $\chi^2$ for
Hessian method 2 as compared to the offset method is $\Delta \chi^2 = 283$.
The results of Hessian method 2 represent a fit with an unacceptably large
value of $\chi^2$ 
when judged in this conventional way. 
\begin{table}
\begin{center}
\begin{tabular}{|l|c|c|c|}
\hline
Experiment & data points& $\chi^2$/data point& $\chi^2$/data point \\
           &            & Hessian method 2 & offset method \\

\hline
 ZEUS96/97 & 242 & 1.37 & 0.83\\
 BCDMS p   & 305 & 0.95 & 0.89 \\ 
 NMC p     & 218 & 1.50 & 1.26\\
 NMC D     & 218 & 1.15 & 0.96\\
 E665 D    &  47   & 0.97 & 0.94 \\
 E665 p    &  47   & 1.17 & 1.16 \\
 CCFRxF3   &  57   & 0.99 & 0.39 \\
 NMC D/p    & 129   & 0.97& 0.93 \\
\hline
\end{tabular}
\caption{Table of $\chi^2$ calculated by adding systematic and statistical 
errors in quadrature for the theoretical parameters determined by the 
offset method and Hessian method 2}
\label{tab:chibab}
\end{center}
\end{table}

\subsection{Parameter evaluation and Hypothesis Testing}

To appreciate the signficance of the difference in $\chi^2$ between various
fits, the distinction between
the $\chi^2$ changes appropriate for parameter estimation and for 
hypothesis testing should be considered. 
Assuming that the experimental uncertainties which contribute have Gaussian
distributions, errors on theoretical 
parameters which are fitted within a fixed theoretical framework
are derived from the criterion for `parameter estimation' 
$\chi^2 \to \chi^2_{min} + 1$. However the goodness of fit of a theoretical
hypothesis is judged on `hypothesis testing' criterion, such that
its $\chi^2$ should be approximately in the range $N \pm \surd (2N)$,
where $N$ is the number of degrees of freedom. 

Fitting DIS data for PDF parameters and $\alpha_s(M_Z^2)$ 
is not a clean situation
of either parameter estimation or hypothesis testing. 
Within the theoretical framework of leading-twist-NLO QCD, 
many model inputs such as the form of the PDF
 parameterisations, the values of cuts, the value of $Q^2_0$, the
data sets used in the fit, etc., can be varied.
These represent different hypotheses and they are accepted, 
provided the fit $\chi^2$ fall within the hypothesis-testing criterion. The
theoretical 
parameters obtained for these different model hypotheses can differ from those
obtained in the standard fit by more than their errors as 
evaluated using the parameter-estimation criterion. In this case the model
uncertainty on the parameters exceeds the estimate of the total experimental 
error. This does not happen for the
offset method in which the uncorrelated experimental
errors evaluated by the parameter estimation criterion are augmented by the 
contribution of the correlated experimental systematic
uncertainties as explained in Section~\ref{sec:errors}. 
The shifts in theoretical parameter values for the different model hypotheses 
are found to be well within the total
conservative experimental error estimates. 
However this is no longer the case when the
fit is performed using Hessian method 2. In this case the shifts in 
theoretical parameter values for the different model hypotheses 
are outside the experimental error estimates. Since our purpose is to estimate
errors on the PDF parameters and $\alpha_s(M_Z^2)$ within a general 
theoretical 
framework, rather than as specific to particular model choices within this
framework, this is an issue which must be addressed.

The CTEQ collaboration~\cite{10,12} have considered 
this problem.
They consider that $\chi^2 \to \chi^2 + 1$ is not a reasonable tolerance
 on a global fit to $\sim 1200$ data points from diverse sources, with 
theoretical and model uncertainties which are hard to quantify and experimental
uncertainties which may not be Gaussian distributed.
They have tried to formulate criteria for a more reasonable setting of the 
tolerance $T$, such that $\chi^2 \to \chi^2 + T^2$ becomes the variation 
on the basis of which errors on parameters are calculated. 
In setting this 
tolerance they have considered that all of the current 
world data sets must be acceptable and compatible at some level,
even if strict statistical criteria are not met, since the conditions
for the application of strict criteria, namely Gaussian error distributions, 
are also not met.
The level of tolerance they suggest is $ T \sim 10 $. Note that 
this is similar to
the hypothesis testing tolerance $T=\surd{\surd (2N)} \sim 7$ for the ZEUS 
fits. The errors for Hessian method 2 have been re-evaluated using the 
tolerance $T=7$ and, for $\alpha_s(M_Z^2)$, the result 
$\alpha_s(M_Z^2) = 0.1120 \pm 0.0033$ 
is obtained. The error is now 
remarkably close to the error estimate of the offset method 
 performed under the same conditions. This is also the case for 
the errors on all of the PDF parameters. 

Thus the offset method and the Hessian method with an augmented
tolerance $T=\surd{\surd (2N)}$ give similar conservative error
estimates. In choosing between these methods there are some additional
considerations. 

In the Hessian method it is necessary to check that data 
points are not shifted far outside their one standard deviation errors. 
When the ZEUS fits are done by Hessian method 2 some of the systematic 
shifts for the 10 classes of systematic uncertainty of the ZEUS data move by
$\sim \pm 1.4$ standard deviations. There is no single kinematic
region responsible for these shifts which could be excluded to reduce this
effect. Whereas these shifts are not very large,  
they differ significantly from the systematic shifts to ZEUS data
determined in the CTEQ fit. The choice of data sets
included in the ZEUS fit also changes the values of these systematic shifts 
and making different model
assumptions in the fits also produces somewhat different systematic shifts. 
It seems unreasonable to let variations in the model, or the choice of data 
included in the fit, change the best estimate
of the central value of the data points.

In summary, the offset method has been selected for several reasons.
Firstly, because its fit results make theoretical predictions which are 
as close to the central values of the published data points as possible. 
The selection of data sets included in the fit or superficial changes 
to the model are not allowed to change 
the best estimate of the central value of the data points. 
Secondly, because its error estimates are equivalent to those of a method 
which does not assume that experimental
systematic uncertainties are Gaussian distributed. Thirdly, because its
results produce an acceptable
$\chi^2$ when re-evaluated conventionally by adding systematic and
statistical errors in quadrature.
Fourthly, because its conservative error estimates 
take account of the fact that the purpose is to estimate
errors on the PDF parameters and $\alpha_s(M_Z^2)$ within a general 
theoretical framework not specific to particular model choices. 
Quantitatively the error 
estimates of the offset method correspond to those which would be obtained
using the more generous tolerance of the `hypothesis testing' criterion
in the more statistically rigorous Hessian methods.

\section{CONCLUSION}

The NLO DGLAP QCD formalism has been used to fit ZEUS data  
and fixed-target data in the kinematic region,
$Q^2>2.5$GeV$^2$, $6.3\times10^{-5}<x<0.65$ 
and $W^2 > 20$GeV$^2$.
The parton distribution functions for the $u$ and $d$
valence quarks, the gluon and the total sea have been determined and the 
results are compatible with those of MRST2001 and CTEQ6.
The ZEUS data are crucial
in determining the gluon and the sea distributions. 

Full account has been taken of correlated 
experimental systematic uncertainties for all experiments.
The resulting experimental uncertainties on the parton distribution functions 
have been evaluated conservatively, such that the model uncertainty, within the
framework of leading-twist next-to-leading-order QCD, is negligible by 
comparison. Hence the error bands on the PDFs resulting from these 
fits represent reasonable estimates of the total uncertainties within 
this theoretical framework.

\vskip1cm
\noindent

\section*{ACKNOWLEDGEMENTS}

I would like to thank the conference organisers  
L.Lyons, M.Whalley, J.Stirling and my
colleagues on the ZEUS experiment.

\end{document}